\documentclass[]{spie} 

\usepackage{amsmath,amsfonts,amssymb}
\usepackage{graphicx}
\usepackage[colorlinks=true, allcolors=blue]{hyperref}

\title{Training telescope operators and support astronomers at Paranal}

\author[a]{Henri M. J. Boffin}
\author[a]{Dimitri A. Gadotti}
\author[a]{Joe Anderson}
\author[a]{Andres Pino}
\author[a]{Willem-Jan de Wit}
\author[a]{Julien H. V. Girard}

\affil[a]{European Southern Observatory, Chile}

\begin{document} 
\maketitle

\begin{abstract}
The operations model of the Paranal Observatory relies on the work of efficient staff to carry out all the daytime and nighttime tasks. This is highly dependent on adequate training. The Paranal Science Operations department (PSO) has a training group that devises a well-defined and continuously evolving training plan for new staff, in addition to broadening and reinforcing courses for the whole department. This paper presents the training activities for and by PSO, including recent astronomical and quality control training for operators, as well as adaptive optics and interferometry training of all staff. We also present some future plans.
\end{abstract}

\keywords{ESO, VLT, quality control, observatory, operations, training, adaptive optics, interferometry}

\section{Paranal Science Operations}
\label{sec:intro} 

\begin{figure} [htbp]
   \begin{center}
   \begin{tabular}{c} 
   \includegraphics[width=15cm]{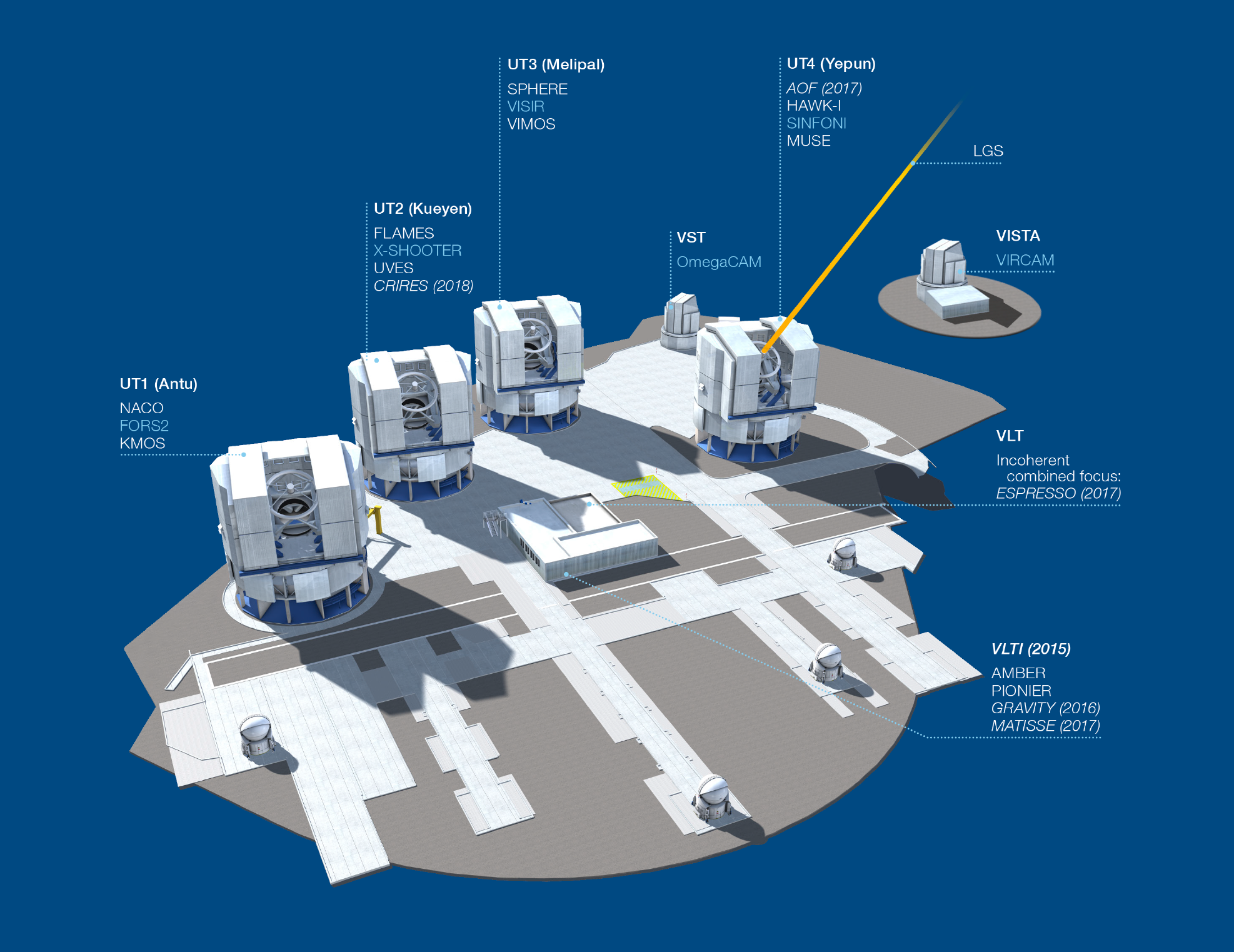}
   \end{tabular}
   \end{center}
   \caption[example] 
   { \label{fig:vlt} 
The Paranal Observatory telescopes and instruments. Instruments listed in blue are at the Cassegrain focii of the telescopes. Instruments listed in italics are not yet installed. Credit: ESO.
}
   \end{figure}

The Very Large Telescope (VLT) at Cerro Paranal in Chile is the European Southern Observatory's (ESO) premier site for observations in the visible and infrared light. Starting routine operations in 1999, it represented at the time the  largest  single
investment  in  ground-based  astronomy ever  made  by  the  European  community. 
At Paranal, ESO operates the four 8.2-m Unit Telescopes (UTs), each equipped with three instruments covering a wide range in wavelength, as well as a variety of technology, from imagers, low- and high-resolution spectrographs, multiplex and integral field spectrographs, to polarimeters. Some of the instruments use adaptive optics technologies to increase their resolution. In addition, the VLT  offers  the possibility of combining the light from the four UTs to work as an interferometer, the Very Large Telescope Interferometer (VLTI), with its own suite of instruments, providing imagery and spectroscopy at the milliarcsecond level and soon, astrometry at 10 microarcsecond precision. In addition to the 8.2-metre diameter telescopes the VLTI is complemented by four Auxiliary Telescopes (AT) of 1.8-metre diameter to improve its imaging capabilities and enable full nighttime use on a year-round basis.
Two telescopes for imaging surveys are also in operation at Paranal, the VLT Survey Telescope (VST, 2.6-metre diameter) for the visible, and the Visible and Infrared Survey Telescope for Astronomy (VISTA, 4.1-metre) for the infrared.
The comprehensive ensemble of telescopes and instruments available in Paranal is depicted in Fig.~\ref{fig:vlt}.

Each year, between 1800 and 2000 proposals are made for the use of ESO telescopes. This leads to ESO being the most productive astronomical observatory in the world, which annually results in many peer-reviewed publications: in 2015 alone, over 950 refereed papers based on ESO data were published. Moreover, research articles based on VLT data are in the mean quoted twice as often as the average. The very high efficiency of ESO's ``science machine'' is not due to chance, but the outcome of a careful operational model, which encompasses the full cycle, from observing proposal preparation to planning and executing the observations, providing data reduction pipelines, checking and guaranteeing data quality, and finally making  the data available to the whole astronomical community through a science archive. 

The operational model of the VLT rests on two possible modes: visitor (or classical) and service. In visitor mode,  the astronomer travels to Paranal to execute their observations. In  service mode, however, observations are queued and executed taking into account, in real time, their priorities and the atmospheric and astronomical conditions. In both cases, observations are done by the Paranal Science Operations (PSO) staff. This is obvious for the service mode, but also necessary for the visitor mode given the very complicated nature of the instruments that cannot be easily taught to the visiting astronomers.

PSO staff is there to 
support observing operations in both visitor (VM) and service mode (SM) in
Paranal. The tasks to be performed include the support of visitor astronomers, the short-term (flexible) scheduling of queue observations, the calibration and monitoring of the
instruments, and the assessment of the scientific quality of the astronomical data, with the ultimate goal of optimising the scientific output of this world leading astronomical facility.

The PSO department consists of about 65 staff\cite{Dumas}, composed of staff astronomers (who do 105 to 135 nights of duties), post-doctoral fellows (80 nights) and telescopes and instrument operators (TIOs). Obviously, this group did not exist on
April  1,  1998, i.e. before First Light of the first UT.  Six  years  later,  it  counted  56
members, a number that thus increased, in parallel to the larger numbers of systems available. 

During the early years of operations, the recruitment and turnover of staff was particularly high. For example, during the first 6 years on average more than 12 PSO staff were recruited each year\cite{Mathys}. While these numbers have now plateaued to lower levels, there is still considerable turnover of staff: in the last 3 years 9 staff astronomers and 3 TIOs have been hired.
Moreover, the post-doctoral fellowships are three-year contracts\footnote{ESO Fellows in Chile receive a $4^{\rm th}$ year which they may choose to spend at ESO Chile, in which case they would have no or reduced duties at the Observatory, but they can also choose to go to a Chilean institution or in a research group in any of the ESO member states.}, so by definition, this leads to a large turnover. This implies that there is a continuous need for a
considerable  training  effort. Moreover, not only is there a need to train the new members of the department, but one also needs to deepen the training of everyone. This article presents how this is done.

\section{Training: General concept}
According to the 
Collins Concise English Dictionary, training is 
``the process of bringing a person to an agreed standard of proficiency by practice and instruction.'' This is exactly what is done for all new members of the PSO department, whether staff astronomer, post-doctoral fellow, or TIO, as well as when a current staff is given more responsibilities. Indeed, astronomers or TIOs are initially assigned to one UT or to the VLTI, but after a year or so, depending on capabilities and willingness, astronomers can be assigned to another UT. Moreover, more experienced astronomers will become day- and night-shift coordinators, which require additional training. Obviously the training will be different for astronomers and for TIOs.

 \begin{figure} [htbp]
   \begin{center}
   \includegraphics[width=15cm]{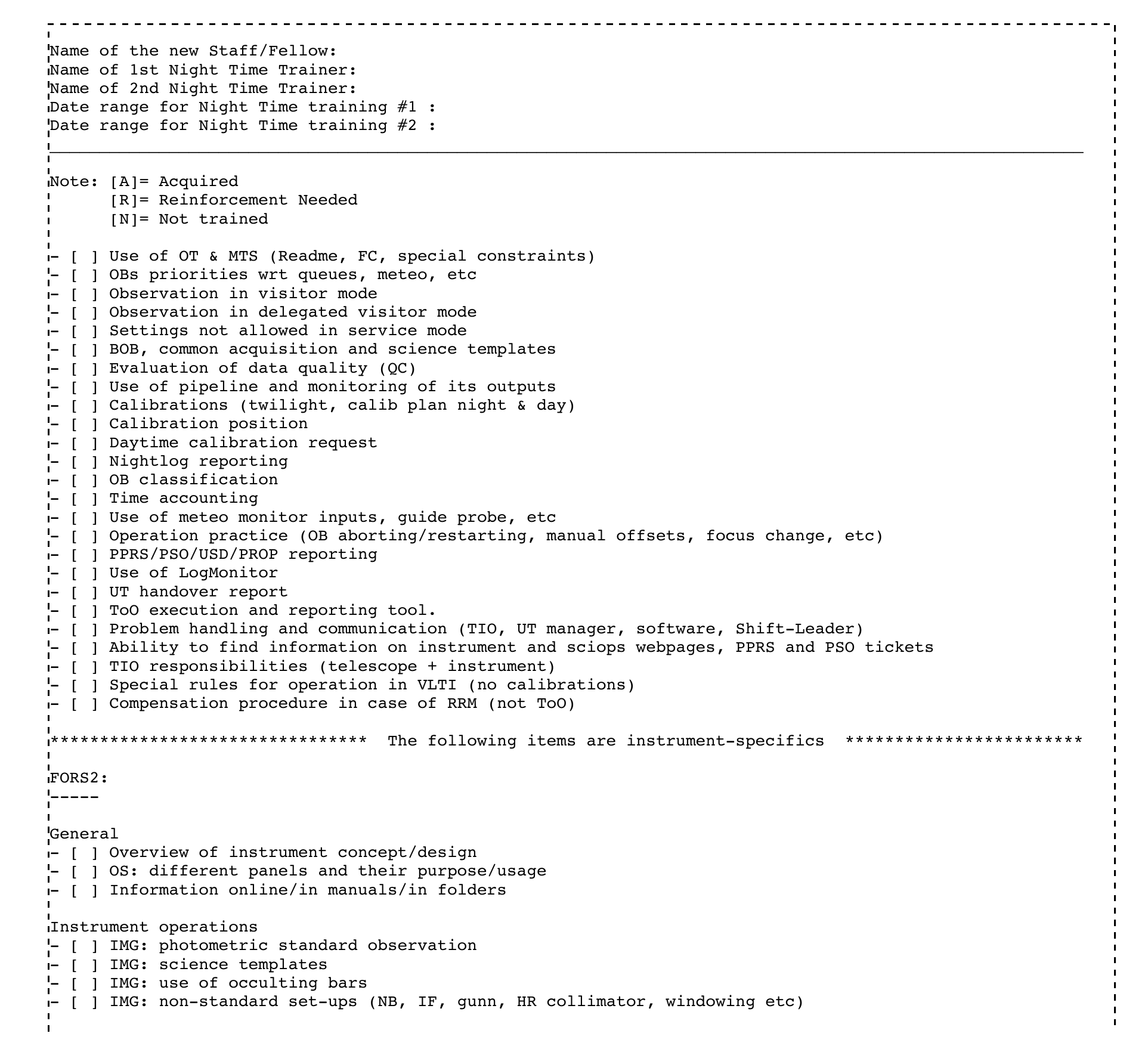}
   \end{center}
   \caption[example] 
   { \label{fig:certif} 
Example of a training certification checklist, in this case, the first part of the checklist to certificate solo operations on UT1.}
   \end{figure}

In the case of astronomers, the initial training consists normally of four shifts of 6--10 days/nights at the Observatory. This is split in two shifts of daytime operations, followed by two shifts of operations at night.  
The first trainer, normally an experienced astronomer, acts as a mentor to ease the start of the new astronomer at Paranal. At the end of this initial training period, the astronomer is certified (see Fig.~\ref{fig:certif}) for solo operations on one system -- either one UT or the VLTI, i.e. they can routinely operate three instruments and are able to troubleshoot problems, together with the TIO. After one year of regular operations, the astronomer is normally able to train others on the instruments they usually work with. 

A certification matrix, constantly updated, indicates the level of knowledge in the various instruments and systems of all PSO staff, allowing a clear view of the level of knowledge in the department. This is useful when making the schedule, in allocating staff to the telescopes and as trainers.

Since 2013, PSO has introduced the concept of Science Operations 2.0 (SciOps 2.0; see [\citenum{Dumas}]). In this scheme, the astronomers assigned to the night-support of the UTs start their ``day'' of duty in the afternoon instead of sunset time and end a few hours (typically ~3-4 hours) before sunrise. The operations during the rest of the night are supported by the TIO on-duty at the telescope, with the help of the UT astronomer who had started duties at sunset and acts as nighttime shift-coordinator for the rest of the night. The implementation of this SciOps 2.0 scheme required additional, specific training\cite{Dumas}. In particular, in order for astronomers to be  
nighttime shift-coordinator, they had to be trained on
the high-level operations of all UT instruments (observing modes, and instrument calibration and quality control plans). Similarly, TIOs received a thorough reinforcement of their instrument operations training in order to be capable of carrying out ``alone'', during the last hours of the night (i.e. under the sole supervision of the nighttime shift- coordinator), the execution of SM and VM science programs. 

\section{TIO training}
\begin{figure} [htbp]
   \begin{center}
   \includegraphics[width=12cm]{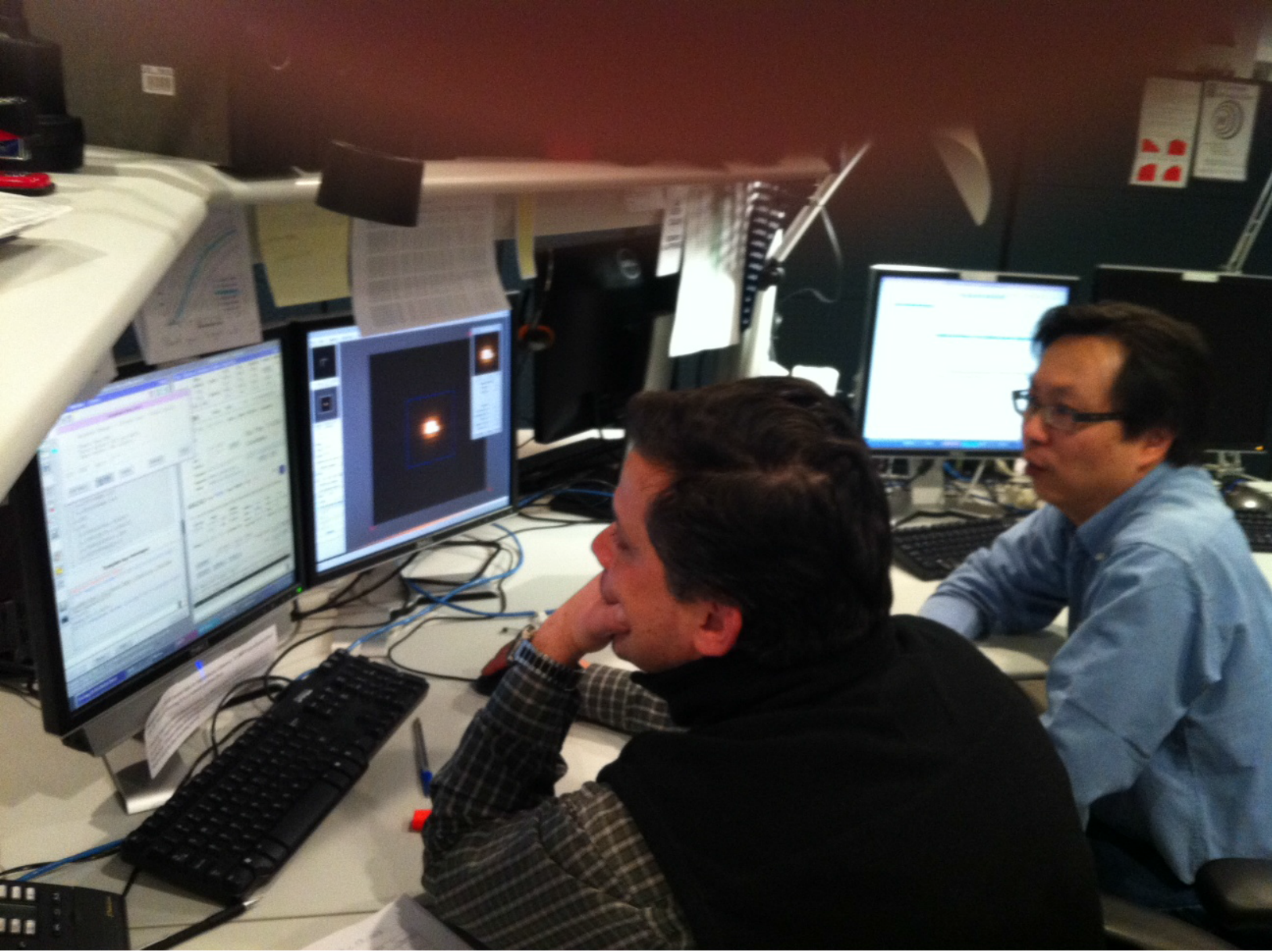} 
   
   \end{center}
   \caption 
   { \label{fig:tio} 
TIO Carlos La Fuente during a training session at UT4, with PSO staff astronomer George Hau (right).}
   \end{figure}

The Telescopes and Instruments Operators Group consists of twenty-four interdisciplinary professionals chosen from different areas of technology: Engineers and Technicians (Physics, Electronic, Electric, Control, Automation and Computing) as well as former Airline Pilots, Air Traffic Controller, Navy officer, Satellite Operator, Antennas Array Operator, etc. Fifteen of them are working exclusively for nighttime operations, while the other nine, called Operations Specialists, share their duties between nighttime and daytime PSO duties.
The activities and level of involvement of the Telescopes and Instruments Operators group during nighttime and daytime has been evolving over the years as per the different operation plans implemented.
For nighttime operations, in order to fulfil the requirements of the current Operations Plan, PSO implemented a standard training system in order to involve all TIOs in the autonomous nighttime operations of the telescopes and instruments installed in the Unit Telescopes (Fig.~\ref{fig:tio}), the Survey Telescopes (VISTA and VST), and the VLTI, as well as their auxiliary subsystems (Laser Guide Star, domes, cooling and atmospheric monitoring systems, etc.). Experienced TIOs are also in charge and act additionally as Weather Officer or Safety Coordinator, depending on scheduling requirements.

The basic training process is standardised for all units, and is performed with direct coaching of the astronomers and the most seniors TIOs. A standard certification process was conceived for every instrument, which was prepared by the respective Instrument Scientist\footnote{Instrument Scientists (IS) are  scientific staff members leading an Instrument Operations Team (IOT) whose purpose is to coordinate all technical and operational activities during the lifetime of the instrument, from its integration into the operations of Paranal, until its decommissioning. The IS is responsible for characterising and validating all observing modes for their instruments, as well as suggest and support the implementation of instrument improvements (new modes, upgrades, improved calibrations, etc).}. On the other hand, the TIOs are in charge of creating documentation about the instrument itself from the operator's point of view (``Survival Guides''), including the definition of data quality control, which is finally used as a training guide for newcomers. In addition, Optics and AO courses are an integral part of the training for all the TIO group (see below).

The activities of TIOs in daytime (Operations Specialists) consist on the support of all daytime operation activities (execution of daytime calibrations, calibration completeness and quality check, mask-manufacturing, instrument troubleshooting, etc.), except for those tasks that specifically require astrophysical expertise (e.g., support of visiting astronomers). These activities are shared with the daytime astronomer (DA).

During the day, the TIOs and DAs are in charge of the completeness and quality certification of the data acquired during the previous night for the assigned telescopes. This includes
the calibration frames acquired in the morning, validating also the content of the night report, and delivering the system to the night astronomer at the beginning of the night with the required instrument set-up and health checks performed. Another significant responsibility of the DA and TIOs is to monitor the instruments through the various quality control systems, and investigate possible deviations. DAs and TIOs are in the front-line in case of instrument problems.
For daytime activities, senior astronomers perform a two-week training activity with the TIOs, with a final certification for these particular duties. 

 In addition to the core duties performed, the TIOs are fully immersed in contributing to all PSO Operations Groups (General Operations, VLTI, Training and documentation, UT Teams, etc.) as well as working for department projects. TIOs are helping in the definition of new standards for operations and developing different projects according to their skills. The background and expertise acquired by TIOs allow them to be the first line of problem detection and a strong
interface with Engineering for helping them in its resolution.

The training plan for TIOs is evolving in time as per the responsibilities and operations scheme in place. For these particular reasons, the additional training plan for TIOs is now a cross-training activity among PSO, giving to the operators a more connected background in astronomical observational, basics in astrophysics for different observation modes, illustration of the science done with the different instruments and workshops for standardisation of the quality assessment of the observations between TIOs and Astronomers.

As a result, TIOs have all attended a two-day workshop about optics in general and adaptive optics. 
Similarly a one-day workshop was also organised that presented astronomical concepts, in particular the following points:
\begin{itemize}
\item how is the astronomical signal altered?
\item Calibrating visible light data: how and why?  
\item IR Observations and their calibrations
\end{itemize}
This workshop was given by senior astronomers from PSO and included numerous practical exercises to familiar TIOs with concepts such as signal to noise ratios and wavelength calibrations.

\begin{figure} [htbp]
   \begin{center}
   \begin{tabular}{cc} 
   \includegraphics[width=7.5cm]{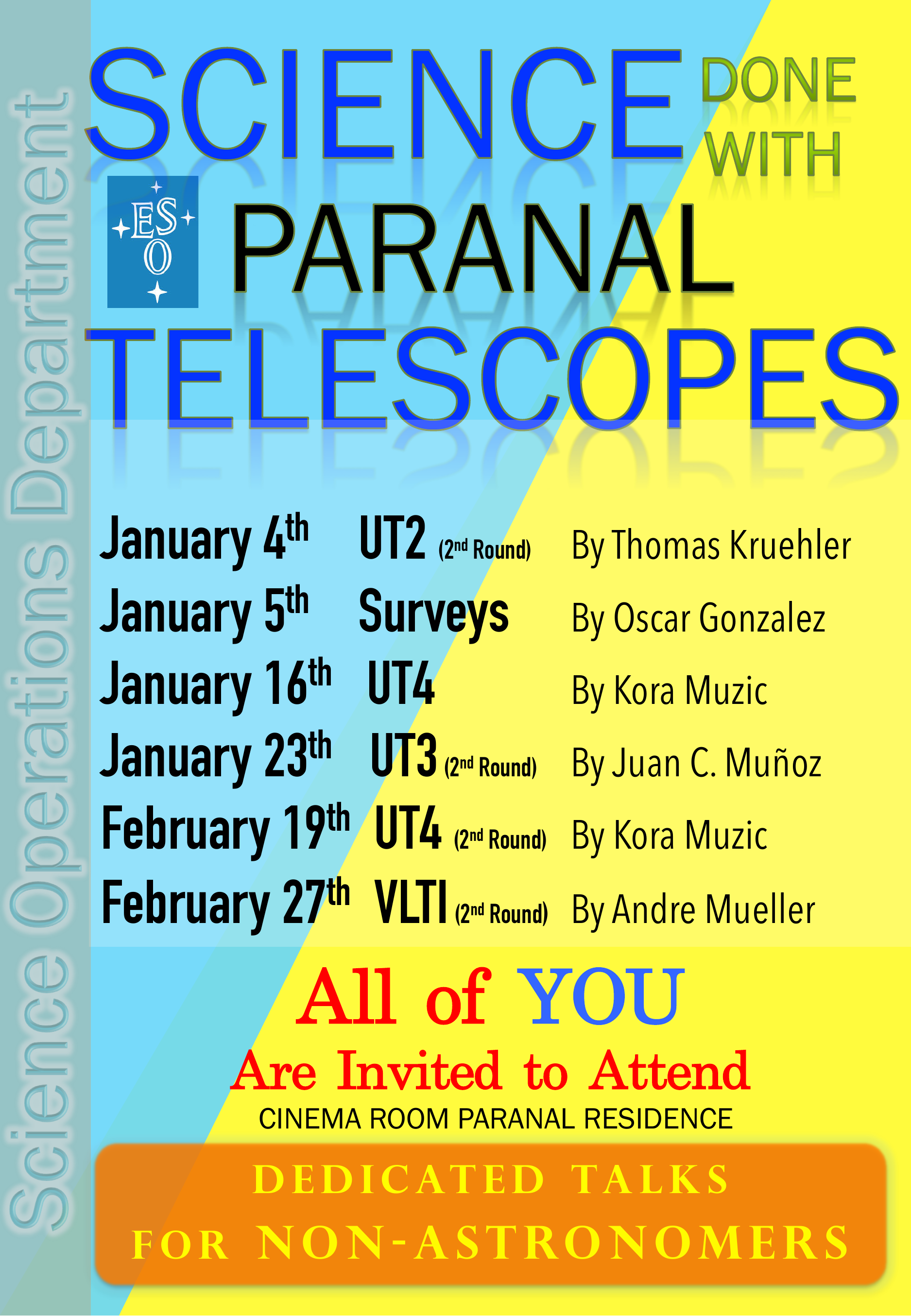}& 
    \includegraphics[width=7.5cm]{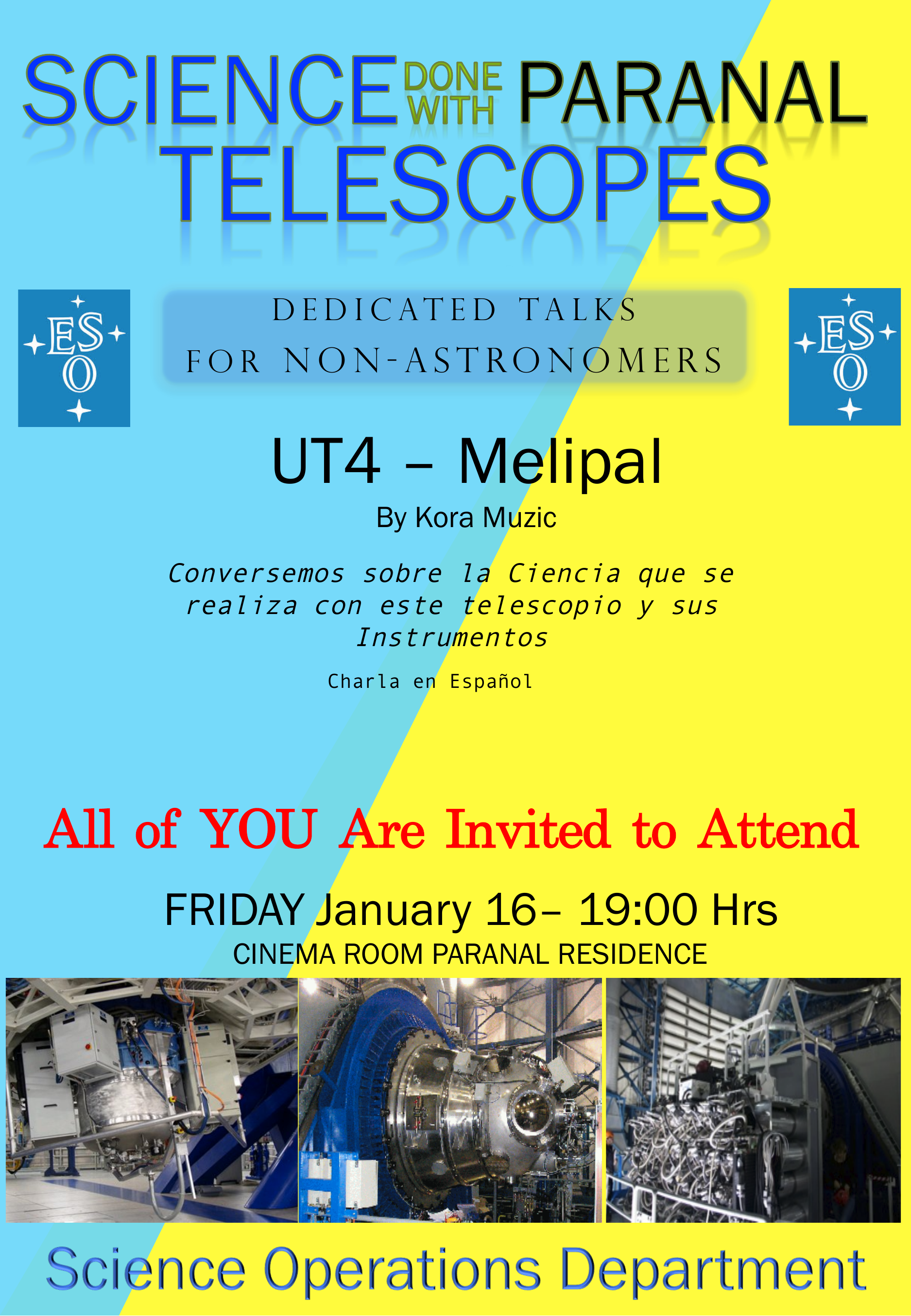}\\
   \end{tabular}
   \end{center}
   \caption[example] 
   { \label{fig:talks} 
Announcements for the series of talks by ESO astronomers for TIOs and Engineers on ``Science done with the Paranal telescopes.''
}
   \end{figure}

In addition, post-doctoral fellows from PSO provided a series of one-hour talks presenting the science done with the various instruments in Paranal 
(Fig.~\ref{fig:talks}). This was aimed not only at TIOs but at all engineers in Paranal, and provided them with a clear understanding of the final objective and results of their day-to-day (or night-to-night) labour.

TIOs also attended, sometimes together with astronomers or engineers, QC0 workshops, adaptive optics and interferometry training, in various sessions. These are now described in more detail.

\section{Quality Control training}
In service mode observations, consistent and accurate evaluation of the data quality 
 is essential in order that the user obtains the data quality they
need for their science. The basic quality control (Quality Control Zero, QC0) is done at the telescope, immediately following the execution of an observing block. Each instrument at Paranal has a set of rules/guidelines to be
used for QC0. An essential goal of the astronomer's and TIO's training is to master QC0. It is essential for efficient operations, as QC0 is the basis for real-time decision-taking at the telescope, and is required to confirm if the observing constraints of the user have been fulfilled.

At Paranal, over the last year, a dedicated QC0 project was started, which 
 consisted of three major stages. The first stage was a discussion among experts of each instrument in order to agree on a well-defined procedure for the quality control at the telescope. For each instrument, the Instrument Scientist is in charge of defining the most appropriate procedure, with the help of the Instrument Fellow and a telescope operator with expertise on the corresponding instrument. As the discussion did not involve every astronomer or operator that provide support to programs using the instrument, it was necessary, once the quality control procedures were consolidated for every instrument, to hold two full-day dedicated workshops where the procedures were presented. The workshops are thus a key stage of the QC0 project, since it allows everyone involved with the operations of an instrument to learn and discuss the procedures, and to resolve any pending questions. Without the workshops, the standardisation of the procedures would be impractical.

During the workshops, aspects of the quality control procedures that are more general and concern more than one instrument were also discussed. Most of these aspects in fact did not actually need to be defined, they already were, but needed to be clarified with a sizeable fraction of the PSO team together. Examples of such aspects are: (i) a reference wavelength for the measure of image quality in spectrographs covering a long wavelength range, and (ii) how to classify imaging OBs with multiple exposures with varying stellar elongations. An essential feature of the workshops was their informal character, that encouraged everyone to participate with comments and raising questions -- the goal was to iron out every doubt, in a lively meeting.

\begin{figure} [htbp]
   \begin{center}
   \includegraphics[width=12.5cm]{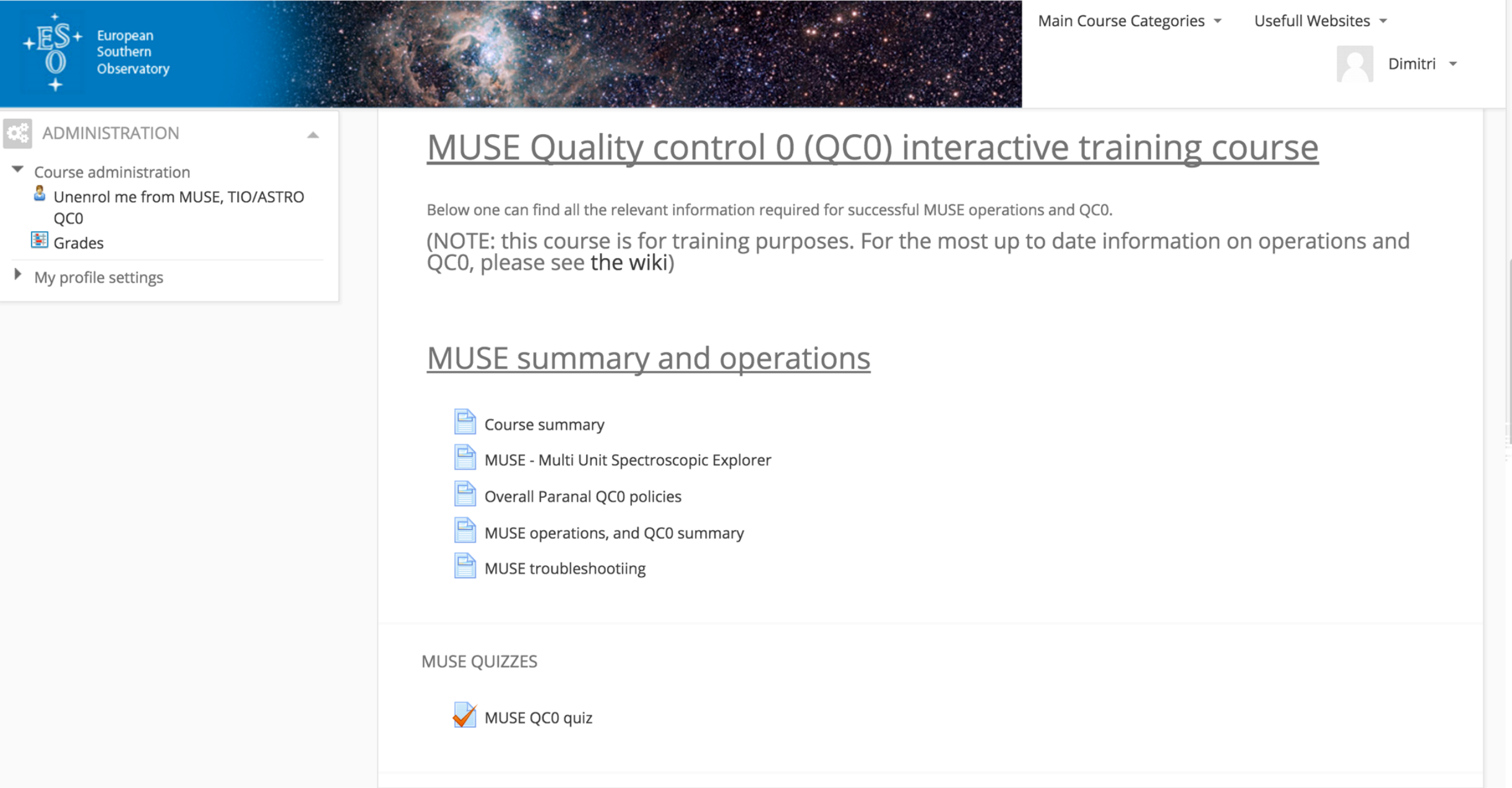}
   \end{center}
   \caption[example] 
   { \label{fig:elarning1} 
Screenshot of one of the courses on the e-learning platform.}
   \end{figure}

\begin{figure} [htbp]
   \begin{center}
    \includegraphics[width=12.5cm]{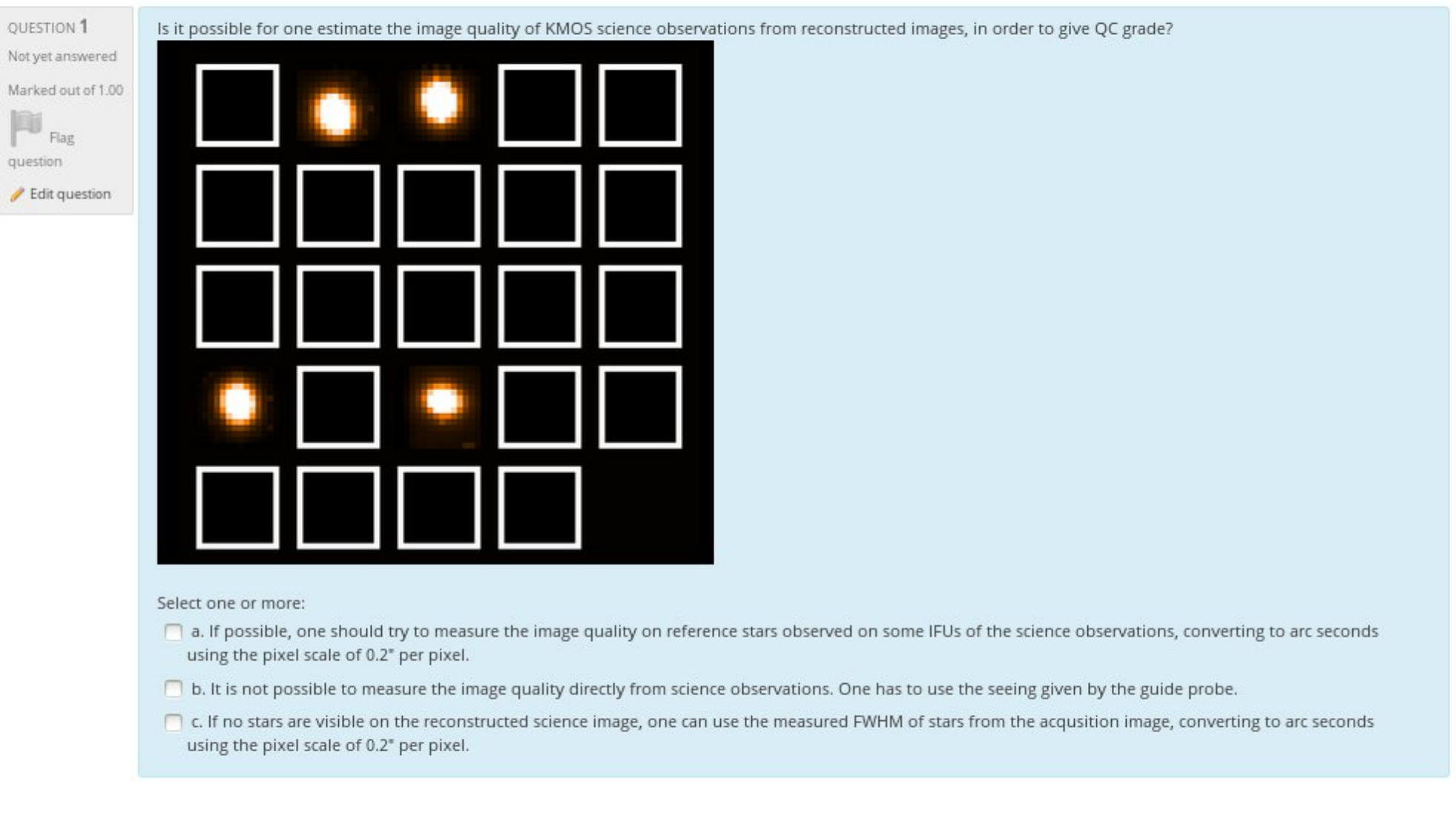}
   \end{center}
   \caption[example] 
   { \label{fig:elarning2} 
An example of one question being part of the quizzes that conclude all courses on the e-learning platform.}
   \end{figure}

The third main stage of the QC0 project was the creation of an e-learning platform with instrument specific courses and quizzes (Figs.~\ref{fig:elarning1} and \ref{fig:elarning2}). These courses contain
all the relevant information needed to do QC0, and then contain multiple-choice quizzes
to test the gained knowledge, and hence provide motivation for further discussion where
inconsistencies/information is not clear.

The e-learning courses use the open source platform {\tt Moodle}\footnote{\url{https://moodle.org/}}. This provides the material
in a user friendly manner, and is easy to implement to contain further questions, new quizzes
etc. Each instrument specific course consists of a set of modules that contains information on 
overall Paranal policies, instrument operations, and instrument specific QC0. The user (Paranal
astronomers or TIOs) is invited to enrol on one of these courses, study the course material,
and finally test the knowledge they have gained (both from the course and in hands-on training 
at the telescope) through answering a series of questions with examples of
different QC0 decisions that will occur during any given night on whichever specific instrument.
Users are then encouraged to discuss any doubts with Instrument Scientists, in order
to further homogenise Paranal QC0 practices.

The e-learning platform proved to be a useful tool for training  in addition to the QC0 practices. While the next training project is now the implementation in the e-learning platform of QC0 practices for the remaining of the instrumentation suite in Paranal -- including VLTI -- it is foreseen that the platform will include training of general procedures. The latter includes issue reporting optimisation/standardisation, improved assessment of quality and quantity of day- and nighttime calibrations.

\section{Adaptive optics training}
All Unit Telescopes of the VLT are equipped with one adaptive optics (AO) system at their Coud\'e focus for VLTI operations. In addition, three UTs feature instruments with dedicated AO modules.  This means that most of the PSO staff (but also the engineers) are now exposed to this technology in its variety of flavours, such as single conjugated AO or extreme AO. In the very near future, UT4 will become a fully adaptive telescope\cite{AOF}, equipped with a deformable secondary mirror, four Laser Guide Stars, and two adaptive optics modules that will provide a corrected light wavefront to the two Nasmyth instruments MUSE and Hawk-I, thereby allowing muti-conjugated AO and tomographic AO. 

Most astronomers have, however, very little knowledge of what is specific about AO, and AO specialists are even rarer. This led PSO to engage in a dedicated training about AO, with the goal to increase general AO knowledge and awareness among all scientific and technical staff in Paranal. It was thus aimed at both the PSO staff and the engineers of the Maintenance and System Engineering department. 

The training consisted in a two-day workshop held at the ESO headquarters in Santiago and repeated twice to allow staff on duties in Paranal to attend. A total 80 persons participated to these sessions. Each day had a specific topic and the various topics covered were the following:

{\bf Day 1: Principles of AO system}
\begin{itemize}
\item Formation of images: basics of geometrical and Fourier Optics, Fraunhofer diffraction, etc.
\item Imaging through turbulence: introduction to optical aberrations
\item Wavefront sensing and correction 
\end{itemize}

{\bf Day 2: Paranal AO instruments}
\begin{itemize}
\item AO Zoology, which AO for which science cases?
\item High Contrast AO: from NACO to SPHERE
\item MACAO and their applications in Paranal 
\item The AOF as a facility (including lasers, DSM, AO modules)
\end{itemize}

The talks and exercises were given by the AO astronomers from the PSO department as well as by engineers who deal with the AO instruments in Paranal. It is the hope that after this workshop the trainees know why each instrument uses AO, how it works, and what are the differences between the various AO flavours. Hopefully, as a result, Paranal staff will be even more motivated about working with such systems.

\section{VLTI training}
As a rather recent addition to the astronomer's toolbox at ESO, the
technique of long baseline optical interferometry is often considered
complex and sometimes even arcane. This perception is related to at
least two factors, {\it viz.} the hardware complexity required to
generate interferograms from spatially separated telescopes, and the
non-intuitive character of the generated data, that is, the interferograms
containing the fringe patterns. Over the years, and since the start of
VLTI operations, the interferometry community in collaboration with
ESO has organised VLTI schools with the aim to remove the doubts
potential users could have had and at the same time enlarge the user
base. The first one of these was organised in Les Houches (France) in 
2002\cite{vlti} and the eighth version took place in September 2015\footnote{See \url{http://www.astro.uni-koeln.de/vltischool2015}}. 

This example shows the continuous need for interferometry training at
large and therefore necessarily within PSO as
well. The underlying reason is the following. The VLTI is the first
optical interferometry that is offered as part of a user facility with
the goal to do astronomy; the VLTI was not conceived as an
interferometric experiment that uses astronomical observations as mere
tests. Apart from users educated at one of the main optical
interferometry centres in Europe (for example OCA Nice, MPIfR Bonn,
IPAG Grenoble), optical interferometry is often not found to be part
of the standard training of astronomers, let alone for telescope
operators.  Training of PSO staff in optical interferometry is 
therefore crucial to ensure efficient operations and the generation 
of quality interferometric data.

To achieve these goals, we implemented a training policy making a
distinction between the needs for TIOs and the needs for
astronomers.

  \begin{figure} [htbp]
   \begin{center}
   \begin{tabular}{cc} 
   \includegraphics[width=7.5cm]{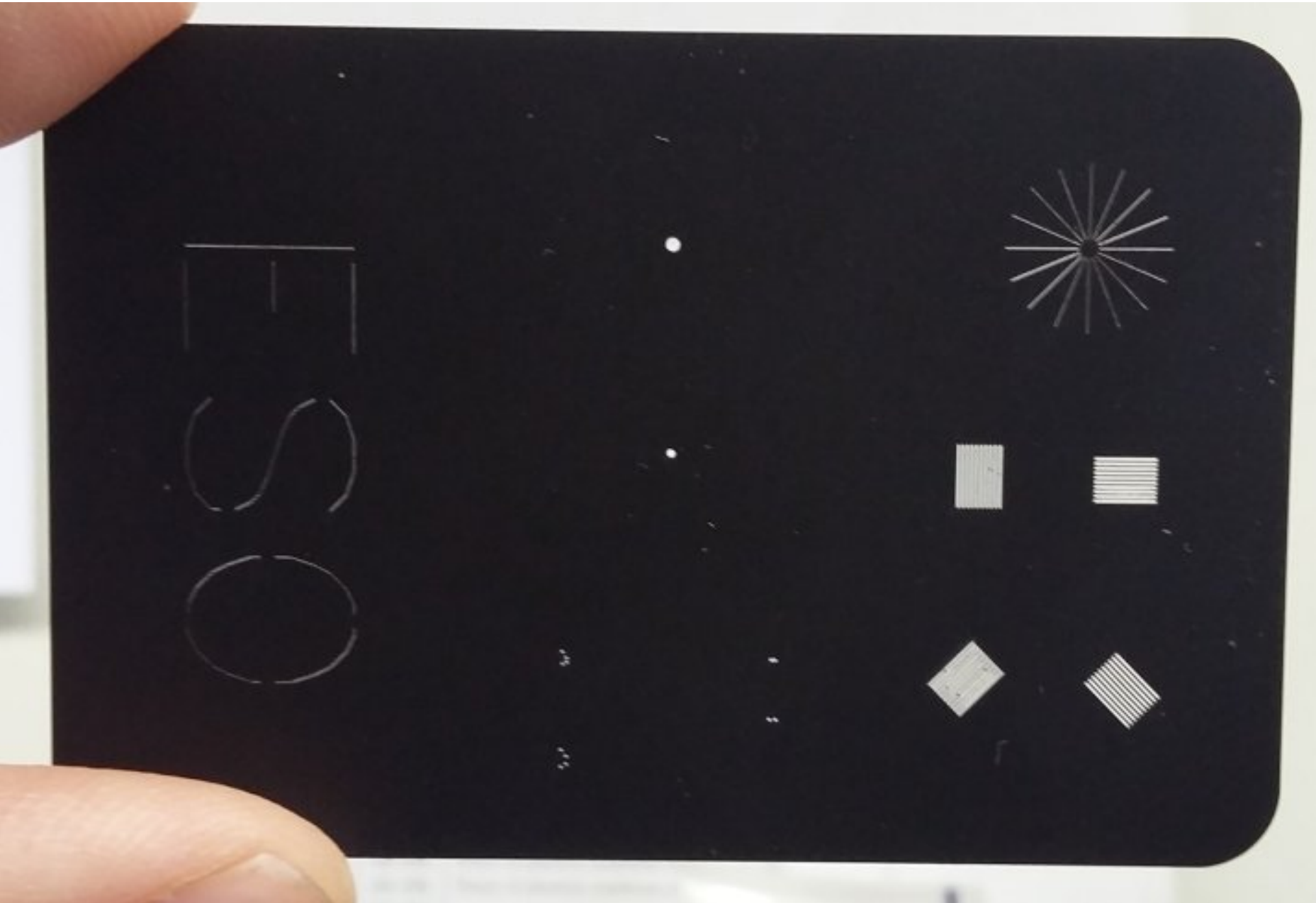} &
   \includegraphics[width=5cm,angle=90]{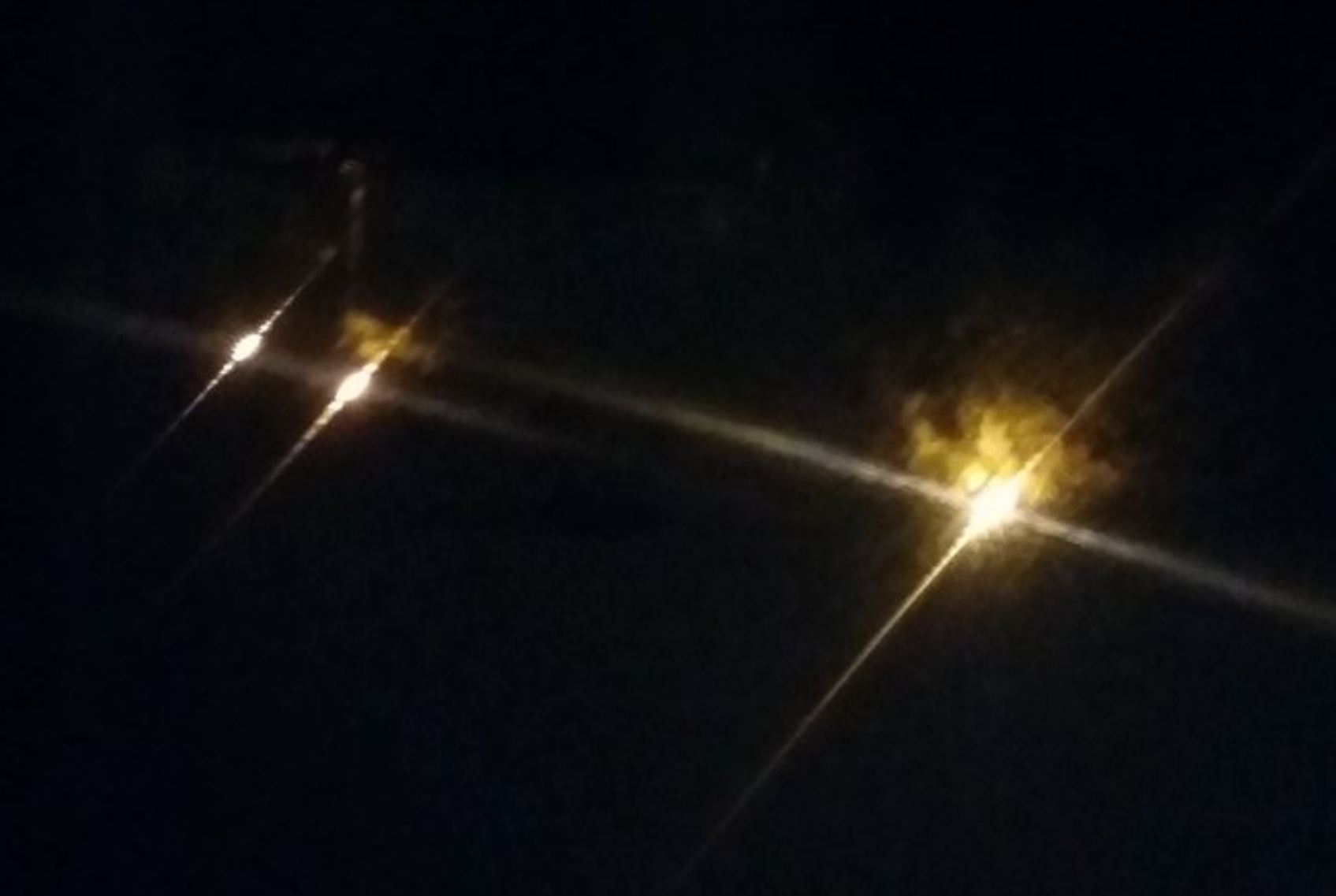}\\
      \end{tabular}
   \end{center}
   \caption 
   { \label{fig:vlti} 
{\it Right:} A combination of apertures punched in a VIMOS mask designed to do hands-on interferometric experiments that clarify the
principle of stellar interferometry. {\it Left:} The diffracted image
created by street lights through two different apertures.}
   \end{figure}

\subsection{Interferometry training for TIOs}
A dedicated workshop
was developed to educate the basics of optical interferometry for
TIOs. Within the VLTI operational scheme, the TIOs are responsible for
the delay-line system and the light-path up to the instrument. As a
result, they work every night with the VLTI hardware and have a good
grasp of the effects of the hardware on the generated data. The workshop
focused therefore on the physical principles of stellar interferometry
and related concepts (spatial coherence, spatial frequencies, Fourier
transform, Van Cittert-Zernike theorem), various detection technique current
in optical interferometry (Michelson and Fizeau, image plane vs. pupil
plane), specific terminology ((z)OPD, ABCD, coherence length, piston)
and an overview of VLTI science.  The workshop consisted of lectures
(given by VLTI astronomers), hands-on backyard stellar interferometry
experiments (Fig.~\ref{fig:vlti}), and problem solving sessions. 

\subsection{Interferometry training for shift coordinators}
All PSO staff that function as
Paranal shift coordinator (SC, i.e. a role that entails coordination of
science operations activities, the representative of the head of PSO
and the interface between the engineering department and PSO) are
required to follow two nights of training of VLTI operations as if
they were night astronomers. The trainer is the VLTI night astronomer. In
addition to reviewing general stellar interferometry concepts, the SC
is made aware of the hardware requirement of VLTI operations
(e.g. telescope relocation, the need for VLTI subsystems like IRIS and the
delay-lines, the FINITO fringe tracker, and the interferometric instruments
 AMBER, PIONIER and Gravity) and the quality control of the
generated data via transfer functions. Additionally, awareness is
created of the differences between operating the VLTI with Auxiliary
Telescopes and Unit Telescopes.

In addition, the training platform (e-learning) as implemented for the Paranal 
instruments (see above) will soon be augmented with one dedicated to the VLTI. 

Training of PSO staff at the VLTI and teaching optical
interferometry is thus a continuous activity undertaken by the VLTI
astronomers.  It is hoped that universities incorporate optical
interferometry systematically into their astronomy courses to create
awareness to the future astronomers that the highest angular
resolution obtainable is through stellar interferometry\footnote{And Astrotomography for some specific astrophysical applications - see [\citenum{AstroTomo}] for a review of all possible techniques in high-angular resolution astronomy.}.

\section{Future plans}
The mid-term future of PSO training in Paranal is devoted to software. In collaboration with the software support team, operators and astronomers will be primed on the software used to control the operations of the VLT\cite{Raffi,Wirenstrand}. The goal is not to train PSO in writing software. But it is evident that a more detailed understanding of the VLT software system will facilitate the communication of needs and issues between PSO and the software team, particularly in the case of new modes and instruments.

The second half of the effort on software training is dedicated to teach the Python language to PSO members using in-house expertise. A few Python scripts are already in use in Paranal, e.g., to measure signal to noise ratios automatically and in real time, to measure the Strehl ratios\cite{ABISM}, and to automate the taking of twilight flat-field frames. The goal is to provide more PSO members with the ability to write scripts to improve efficiency and data quality and minimise problems.

\section{Conclusion}
We have briefly presented the various training activities organised for staff of the Paranal Science Operations department. These activities, made mostly using only internal resources, are now going well beyond the initial (and necessary) training of newcomers in the department, and aim at increasing the efficiency and motivation of the people that operate the Paranal telescopes and instruments. It becomes more and more important to provide training on specialised topics such as adaptive optics and interferometry, as well as on modern programming languages. Such training activities will still need to be developed in the future to address new challenges coming with more and more sophisticated instruments. 

 

\pagebreak
\bibliographystyle{spiebib} 

\end{document}